\documentclass[12pt]{iopart}
\eqnobysec
\usepackage{graphicx}
\usepackage{color}

\newcommand{\be}{\begin{equation}}
\newcommand{\ee}{\end{equation}}

\newcommand{\vecp}{{\mathbf p}}
\newcommand{\vecq}{{\mathbf q}}

\newcommand{\x}{{\mathbf x}}

\newcommand{\X}{{\mathbf X}}

\newcommand{\y}{{\mathbf y}}

\newcommand{\J}{{\mathbf J}}

\newcommand{\mH}{{\mathbf H}}

\newcommand{\M}{{\mathbf M}}

\newcommand{\Id}{{\mathbf I}}

\newcommand{\der}{\partial}

\newcommand{\vct}[1]{\ensuremath\mbox{\boldmath$ #1 $}}
\newcommand{\Vxi}{{\vct \xi}}

\newcommand{\z}{\vct z}

\begin{document}

\title{The quantum canonical ensemble in phase space}

\author {Alfredo M. Ozorio de Almeida\footnote{ozorio@cbpf.br}}
\address{Centro Brasileiro de Pesquisas Fisicas,
Rua Xavier Sigaud 150, 22290-180, Rio de Janeiro, R.J., Brazil}

\author{Gert-Ludwig Ingold}
\address{Institut f\"ur Physik, Universit\"at Augsburg, Universit\"atstra{\ss}e 1, D-86135 Augsburg, Germany}

\author{Olivier Brodier}
\address{Institut Denis Poisson, Campus de Grandmont, 
Universit\'e de Tours, 37200 Tours, France}

\begin{abstract}

The density operator for a quantum system in thermal equilibrium with its environment
depends on Planck's constant, as well as the temperature. At high temperatures,
the Weyl representation, that is, the thermal Wigner function, becomes indistinguishable
from the corresponding  classical distribution in phase space, whereas the low temperature limit
singles out the quantum ground state of the system's Hamiltonian. In all regimes,
thermal averages of arbitrary observables are evaluated by integrals, as if the thermal
Wigner function were a classical distribution. 

The extension of the semiclassical approximation for quantum propagators to an imaginary thermal time, 
bridges the complex intervening region between the high and the low temperature limit. 
This leads to a simple quantum correction to the classical high temperature regime, 
irrespective of whether the motion is regular or chaotic. 
A variant of the full semiclassical approximation with a real thermal time, though in a doubled phase space, 
avoids any search for particular trajectories in the evaluation of thermal averages.
The double Hamiltonian substitutes the stable minimum of the original
system's Hamiltonian by a saddle, which eliminates local periodic orbits from the stationary phase 
evaluation of the integrals for the partition function and  thermal averages.

\end{abstract}

\maketitle

\section{Introduction}

The classical canonical ensemble is characterised by the {\it thermal probability density}:
Given the {\it Hamiltonian}, $H(\x)$, for a physical system with $N$ degrees of freedom, 
defined in the $2N$-dimensional {\it phase space} with coordinates 
$\x = (\vecp,\vecq)=(p_1,...,p_N, q_1,...,q_N)$, the probability density is
\be
P_{\beta}(\x) \equiv \frac{\exp\{-\beta H(\x)\}}{\int {\rm d}\x~ \exp\{-\beta H(\x)\}}.
\label{clasdis}
\ee
Overall, the classical motion is assumed to be bounded, with an absolute minimum for the Hamiltonian.
The only free parameter is $\beta=1/\kappa_B T$, where $T$ is the temperature and
$\kappa_B$ is Boltzmann's constant. (Physically, the thermal distribution results through equilibrium
with an external environment, which need not be further invoked here.) Since $1/\beta$ is the half-width 
of the energy distribution, it is evident that $P_{\beta}(\x)$ becomes quite irrelevant 
for the description of the true physical system, if $\kappa_B T$ is lower than the 
ground state energy of the corresponding quantum system. 

The {\it thermal density operator}
\be
\hat{\rho}_\beta \equiv \frac{1}{Z_\beta}~ \e^{-\beta \hat{H}},
\label{rhobeta}
\ee
replaces the thermal probability density in the quantum description of
the canonical ensemble of states in thermal equilibrium;
the {\it partition function} being defined as
\be
Z_\beta\equiv \rm{tr} ~ \e^{-\beta \hat{H}}.
\label{partition}
\ee
This is a mixed state of the {\it eigenstates} $|j\rangle$ of the Hamiltonian $\hat H$, 
which is constrained here only by its energy spectrum, assumed to be discrete, countable and bounded from below.
Choosing the sequence of {\it eigenenergies} $E_j$ to be non-decreasing with the index $j$, 
the spectral decomposition,
\be
\e^{-\beta \hat{H}}= \sum_j  \e^{-\beta E_j}~|j\rangle\langle j|,~~~~ 
Z_\beta= \sum_j  \e^{-\beta E_j},
\label{rhobeta2}
\ee
reveals that at low temperatures $\hat{\rho}_\beta$ has no relation to the classical distribution $P_{\beta}(\x)$.
Instead it is dominated by the ground state $|0\rangle$ and the lowest excited states. 

It is important to note that, even though Planck's constant $\hbar$ does not appear explicitly
in \eref{rhobeta}, it fundamentally affects both the eigenstates
and the density of states. Indeed, it can be an advantage 
to consider the non-normalized thermal operator as a continuation of the unitary evolution operator 
\be
\hat{U}_t = \e^{-it{\hat H}/\hbar}
\ee
for the {\it imaginary thermal time}  $t= -i\theta$ with $\theta = \hbar\beta$.
Thus, by relaxing $\beta$ to be positive or negative, these operators also form a group  
in one-to-one correspondence
with the group of unitary evolution operators. There may be no physical relevance for the group
properties of the entire set of thermal operators, but Feynman \cite{Feyn72} employs their Abelian subgroups, 
which share the same constant Hamiltonian, to construct path integrals for the thermal operators with finite $\beta$.   

Of course, Planck's constant is fixed, but the ratio of $\hbar$ to an appropriate action of the system 
is the standard parameter for semiclassical (SC) approximations to the evolution operator and so it is considered here.
There are several methods for the construction of SC approximations, 
including the stationary phase evaluation of path integrals (see e.g. \cite{Gutzbook}).
It is curious that Feynman never explored these possibilities, neither for the evolution operator,
nor as an approximation to the thermal operator, such as presented in \cite{IngoldLec}.
This extension, our present concern, requires attention to the delicate interaction of both parameters, $\beta$ and $\hbar$.
Indeed, it is somewhat paradoxical that one should consider the {\it thermal time} within the classical Hamiltonian flow itself,
even though $\theta = \hbar\beta$. 

The standard choice for the investigation of the properties of the density operator
is the position representation, that is, the density matrix, as in \cite {Feyn72}.
Even so, it is the Weyl representation, which best spans the deep gulf
between the classical and the quantum regimes.
Observables $\hat O$ are represented by real functions $O(\x)$, which usually equal 
(or closely approximate) the corresponding classical variable, whereas 
Wigner functions $W(\x)$ were conceived for the purpose of evaluating quantum
averages in the same way as probability distributions in the classical phase space \cite{Wigner}.
Thus, the expectation is
\be
\langle \hat{O} \rangle = {\rm tr}~\hat{\rho}~\hat{O} = \int {\rm d}\x ~ W (\x)~O(\x),
\label{average}
\ee 
even though the real function $W(\x)$ generally does assume negative values.
For a pure state $|\psi\rangle$ the density operator is the projector, 
$\hat{\rho} = |\psi\rangle\langle \psi|$, whereas a mixed state
is a superposition of projectors onto orthogonal states, weighed by their probability.
Given that the identity operator, $\hat I$, has the Weyl representation $I(\x)=1$,
so that
\be
\langle \hat{I} \rangle = {\rm tr}~\hat{\rho}~\hat{I} = \int {\rm d}\x ~ W (\x) =1,
\label{normalize}
\ee 
the usual convention for the normalization of the Wigner function differs 
from the Weyl representation of $\hat\rho$: $\rho(\x)= (2\pi\hbar)^N W(\x)$. 

Often, observables correspond closely to classical phase space functions
\footnote{For the Hamiltonians that follow, we will not distinguish the classical from the Weyl functions.}, 
so that the integral in \eref{average} may well wash away 
detailed oscillatory structures of the Wigner function.
But $\hat{O}$ may also stand in for another projector, $|\phi\rangle\langle \phi|$,
so that here $\langle\hat{O}\rangle={\rm Pr}(\phi)$, 
the probability of finding the system in the state $|\phi\rangle$. In particular, 
for $|\phi\rangle = |\X\rangle$, the {\it coherent state} centred on the phase space point $\X$, for which 
\be
W_{\X}(\x) = \frac{1}{(\pi\hbar)^N} \exp \left[-\frac{(\x-\X)^2}{\hbar}\right],
\label{coherents}
\ee
so that one can choose $O(\x)= (2\pi\hbar)^N W_{\X}(\x)$,
the integral for the probability ${\rm Pr}(\X)$ will pinpoint a classically minute 
(possibly negative) region of the Wigner function. 
Last but not least, the Hermitian reflection operator, around the point $\X$, is given 
in terms of the vector operator $\hat{\x} = (\hat{\vecp}, \hat{\vecq})$ as
\be
\hat{R}_\X = \int \frac{{\rm d}\x}{(4\pi\hbar)^N} \exp\left[\frac{i}{\hbar}\x\wedge(\hat{\x}-\X)\right].
\ee
Also known as the parity operator around $\X$, it is a bona fide observable, experimentally measured in quantum optics 
\cite{Bertet02}, and it is the defining operator for the Weyl representation
and the Wigner function \cite{Grossmann, Royer,Report}:
\be
W(\X) \equiv \frac{1}{(\pi\hbar)^N}~ {\rm tr}~\hat{\rho}~\hat{R}_{\X}.
\ee
The fact that the Wigner function itself may be considered as the average of the parametrized observable
$\hat{R}_\x$ exemplifies the quantum richness of this seemingly classical phase space representation. 
 
The subject of our study is the {\it thermal Wigner function}, $W_{\beta}(\x)$, 
the appropriately normalized Weyl representation of the thermal density operator \eref{rhobeta}.
In its spectral decomposition, we then have
\be
W_{\beta}(\x)= \frac{1}{Z_\beta} \sum_j  \e^{-\beta E_j}~ W_j(\x),
\label{thermalW}
\ee
where $W_j(\x)$ is the pure state Wigner function which represents the eigenstate $|j\rangle$
in phase space. In the following section  a schematic overview of the thermal Wigner function 
discusses the extreme regimes: At very low temperatures energy scales are accessed where it depends only on $\hbar$,
whereas one reaches a purely classical dependence on $\beta$ at very high temperatures. 
The objective is to present a full SC theory that
bridges these regimes. It will be shown that the first simple correction to the high temperature limit  
depends on a dimensionless parameter, which combines the product of both basic physical parameters with
a further relevant property of the classical Hamiltonian.

The following section provides a quick overview of the full range of variation of the thermal Wigner function.
Then, in section 3, we exploit the direct course of extending analytically the evolution operator 
within the Weyl representation, i. e. the {\it Weyl propagator} \cite{Ber89,Report}, 
in terms of the {\it imaginary thermal time} . 
This is achieved for quadratic Hamiltonians for which the SC approximation is closed and exact.
Not only does this procedure provide the low temperature limit of the thermal Wigner function
for a generic Hamiltonian with a quadratic minimum, but a local expansion sheds light on 
the high temperature limit of the SC approximation. A further employment of analytical continuation 
in section 4 also allows for explicit formulae for the SC approximation
for a Hamiltonian in its normal form, including the Kerr Hamiltonian as a special case.
It should be noted that the exposition up to this section
already supplies valuable corrections to the classical high energy approximation,
which do not depend on the intricacies of the full SC approximations, which then follow.

We now depart from a previous course of pushing through the SC approximation within
the complexified phase space \cite{BroMalAlm20}.
Our method here is based on the expression of the SC Weyl propagator in a doubled 
real phase space, which results by splitting each classical trajectory in a pair of
half trajectories moving backwards and forwards in time. This is presented in section 5, 
where the ordinary phase space variables, $\x= (\vecp,\vecq)$, play the role of {\it double positions}. 
To this one adds a $(2N)$-dimensional space of {\it double momenta}, $\y$, just as in the previous
SC treatment of non-unitary evolution \cite{OARiBR09,BrOA10}, so that the relevant classical motion
is generated by a double Hamiltonian $I\!\!H(\x,\y)$ constructed from $H(\x)$. In section 6 
the complexification of time is incorporated within an imaginary double momentum, 
following which we retrieve a real double phase space. Here the relevant classical motion 
is generated by a new real double Hamiltonian. 

In section 7 the partition function and the thermal average of an arbitrary observable $\hat{O}$ in \eref{average} 
are then obtained in terms of initial value for the pair of forward and backward trajectories, instead of the implicit 
boundary conditions in standard SC methods. Saddle point approximations for the partition function and expectations
are then considered, pointing to the absence of periodic orbit contributions (that are prevalent for real times)
within the neighbourhood of the minima of the Hamiltonian.

The SC scenario in which the thermal Wigner function effects a {\it pseudo-evolution} in thermal time 
is then discussed in section 8. An appendix presents the SC version of some standard thermodynamic relations.

\section{Overview of the thermal Wigner function and its parameters}

The temperature appears explicitly in the exact expression \eref{thermalW} 
for the thermal Wigner function, whereas Planck's constant $\hbar$
is the fundamental parameter within each of the pure state Wigner functions, as well as in the
density of states. Strictly, a SC limit of the thermal Wigner function
would be the result of inserting a good SC approximation for each eigenstate
in \eref{thermalW}, for arbitrary values of the thermal parameter $\beta$. But it is
the joint dependence on  both parameters $\beta$ and $\hbar$ within a wide scenario 
that colours the various regimes of the thermal Wigner function. Hence, the term
{\it SC approximation} is extended here, {\it lato sensu}, as an adaptation
of the powerful asymptotic methods underlying these approximations in the context where
the product $\hbar\beta$ slyly assumes the role of a thermal time 
even in the underlying classical context.

Even though we will not work with the spectral decomposition of the thermal
Wigner function directly, it does provide a rough guide to the various regimes.
First, let us consider the classical Hamiltonian, $H(\x)$: Here it must have an absolute minimum, 
corresponding to the energy spectrum being bounded from below, and we are free to
place it at the origin of phase space with $H(0)=0$. 
This stable equilibrium has generically the lowest Taylor approximation
\be
H(\x|\mH) \equiv \frac{1}{2}~\x\cdot \mH~ \x~,
\label{quadraticH}
\ee
a real positive quadratic form defined by the Hessian matrix of the Hamiltonian, 
whatever the number of degrees of freedom. It will be important here to consider
(homogeneous) quadratic Hamiltonians as a class on their own, parametrized by the matrix $\mH$.
This generates the class of all linear canonical, i.e. {\it symplectic transformations}.

If $\hbar$ is {\it small enough}, the lowest eigenenergies lie in the quadratic region
and, in the limit of extremely low temperature, the spectral  decomposition \eref{thermalW} is dominated 
by the ground state of the quantized  version of \eref{quadraticH}. The explicit form of this Wigner function is reduced in the 
following section to that for a product of $N$ harmonic oscillators, that is, $ W_{\X=0}(\x)$ in
\eref{coherents}. Increasing $\beta$ beyond this point does not affect the thermal Wigner function,
which is only parametrized by $\hbar$. On the other hand, if a slight increase of temperature
brings in just a few significant states within the spectral decomposition \eref{thermalW},
while still lying within the quadratic region of the Hamiltonian, these will be {\it Fock states}, 
with Wigner functions given by their Gr\"onewold expression \cite{Gronewold46}
\begin{equation}
\label{Fockst}
W_j(\x) = 
\frac{(-1)^j}{\pi \hbar} e^{-\x^2/\hbar}
L_j \left(\frac{2\x^2}{\hbar}\right) \; ,
\end{equation}
where $L_j$ is the $j$'th Laguerre polynomial.

Further decrease of $\beta$ allows states at higher energy, for which the quadratic approximation of $H(\x)$ 
no longer holds, to participate in the expansion \eref{thermalW} for the thermal Wigner function, 
but an integrable approximation via normal forms can still be quite accurate \cite{Arnold,livro}. 
This is the regime where the SC approximation \cite{Ber77} (see also \cite{AlmHan82}) constructs each $W_j(\x)$,
the Wigner function for a (nearly) integrable eigenstate,
in terms of the Airy function, enveloping each Bohr-quantized torus. For even smaller $\beta$,
energies corresponding to a saddle point of $H(\x)$ may be accessed, or in any case, chaotic motion
may well become prevalent, so that the point is reached where no SC approximation
for an individual state is available any longer.

Notwithstanding this basic difficulty, one may still have recourse to Berry's semiclassical
approximation for a {\it microcanonical ensemble of states}, coarse grained 
over an energy window that is classically very narrow, but still contains many eigenstates, 
due to their high semiclassical density \cite{Ber89b}. Then one is freed from the details of the
classical motion, whether integrable, chaotic or mixed, and the basic feature is a single
Airy function that reaches a maximum very near the classical energy shell, decays outside of it
and oscillates inside (reaching into a rich structure of caustics \cite{Report}). 
Crudely, this microcanonical Wigner function can be approximated as a Dirac delta-function 
over the energy shell, which, even so, gives reasonable averages  \eref{average} 
for smooth classical observables.

In any case, a general feature is that each contribution
to the spectral sum \eref{thermalW} hardly affects the thermal Wigner function outside its corresponding 
energy shell. This indicates that the classical approximation, $W_\beta(\x) \approx P_{\beta}(\x)$
given by \eref{clasdis}, is at least an envelope for the thermal Wigner function 
at the high energies accessed by a high temperature. 
But the same result follows from keeping just the first term in the power expansion 
of the exponential operator $\exp\{-\beta \hat{H}\}$ in the Weyl representation.
Such an expansion requires a high temperature, but the full validity of this approximation will be seen
to depend on the smallness of the {\it thermal time} $\theta=\hbar\beta$ and on the local form of $H(\x)$.

\section{Analytic continuation of the Weyl propagator}

The quantum flow $\hat{U}_t$ generated by a Hamiltonian $\hat H$ may be SC-approximated
in terms of the classical canonical flow $\x_- \mapsto \x_+(t)$ generated by the corresponding
classical Hamiltonian $H(\x)$. The {\it Weyl propagator}, its Weyl representation, 
has the SC-approximation \cite{Ber89, Report}
\begin{equation}
U_t(\x) \approx \frac{2^N}{|\det(\Id+ \M_t)|^{1/2}}\>\>
\exp \left[ \frac{i}{\hbar}(S_t(\x))\right],
\label{Uweyl}
\end{equation}
in the simplest instance.
The geometric part of the {\it centre (or Weyl) action} $S_t(\x)$ is just the symplectic area between
a trajectory arc centred on $\x$, that is, with endpoints $\x_\pm$  satisfying $2\x={\x}_+ +{\x}_-$,
and its geometric {\it chord} $\Vxi={\x}_+ -{\x}_-$, as shown in Fig.1.
From this, one subtracts $Et$, where $E=H(\x_-)=H(\x_+)$ (and generally $E\neq H(\x)$).
For short times, it is guaranteed that there exists a single trajectory satisfying the 
boundary condition. Eventually, the amplitude may become singular (at a caustic); 
beyond this, there is more than one chord for each centre, 
so that the propagator becomes a sum over terms like \eref{Uweyl}. Each term then has
an extra {\it Maslov phase} in the exponent, just a multiple of $\pi$ in the present context \cite{OAI}.

The trajectory $\tilde{\x}(t',\x_-)$, such that for the full interval $\x_+ = \tilde{\x}(t,\x_-)$, 
is generated by Hamilton's equations
\be
\dot{\x} = \J \frac{\der H}{\der \x}, ~ {\rm or}~~ 
\dot{\vecp} = -\frac{\der H}{\der \vecq} ~{\rm and} ~ \dot{\vecq} = \frac{\der H}{\der \vecp},
\ee
where the standard symplectic matrix in Hamilton's equations is
\be
\J = \left(
\begin{array}{cc}
     0 & -1 \\
     1 & 0 
\end{array}
\right) 
\ee 
in terms of $(\vecp,\vecq)$ blocks. Then the explicit expression for the centre action is
\be
S_t(\x) = \int_0^t \tilde{\vecp}(t',\x_-)\cdot \dot{\tilde{\vecq}}(t',\x_-)~ {\rm d}t' 
-\vecp \cdot (\vecq_+ - \vecq_-) - t H(\x_-).
\label{St1}
\ee
The classical role of the centre action is that of a {\it generating function} 
of the canonical transformation ${\x}^- \mapsto {\x}^+$,  indirectly through \cite{Report}
\begin{equation}
\Vxi = -\J \frac{\der S_t}{\der \x}~~ {\rm and} \>\> {\x}_+ = \x + \frac{\Vxi}{2},
\>\> \x_-  = \x - \frac{\Vxi}{2}.
\label{centran}
\end{equation}

The linear approximation of this transformation near the $\x$-centred trajectory is defined
by the {\it symplectic monodromy matrix} $\M_t$. This has the Cayley parametrization \cite{Report}:
\begin{equation}
\M_t = [\Id + \J\mathbf{B}_t]^{-1} [\Id - \J\mathbf{B}_t],
\label{Cayley}
\end{equation}
in terms of the symmetric Hessian matrix
\be
\mathbf{B}_t(\x) \equiv \frac{\der^2 S_t(\x)}{\der \x^2}
\label{Hessian}
\ee
and the identity matrix $\Id$. This allows for the alternative form of the SC propagator \eref{Uweyl} as \cite{Report}
\begin{equation}
U_t(\x) \approx {|\det(\Id\pm\J \mathbf{B}_t(\x))|^{1/2}}\>\>
\exp \left[ \frac{i}{\hbar}(S_t(\x))\right].
\label{Uweyl2}
\end{equation}

A fundamental property of the centre action is that it is an odd function of time,
that is, $S_{-t}(\x)=-S_t(\x)$, since $S_0=0$ and  the exchange $\x_+\leftrightarrow \x_-$ 
merely reverses the sign of $\Vxi$ in \eref{centran}. It follows from the first of these equations 
that, introducing the {\it wedge product} $\Vxi \wedge \x = (\J\Vxi) \cdot \x$,
the local {\it plane wave approximation} of the action is 
\be
S_t(\x')\approx S_t(\x) + \Vxi\wedge(\x'-\x) ,
\ee
so that $\J\Vxi(\x)/\hbar$ is the {\it local wave vector} in phase space of the Weyl propagator. 
The full expansion of the centre action as a power series in time has only odd terms 
and hence the complexification $S_{it}(\x)$ is a purely imaginary function for all real centres, $\x$.

\begin{figure}
\centering
\includegraphics[width=.8\linewidth]{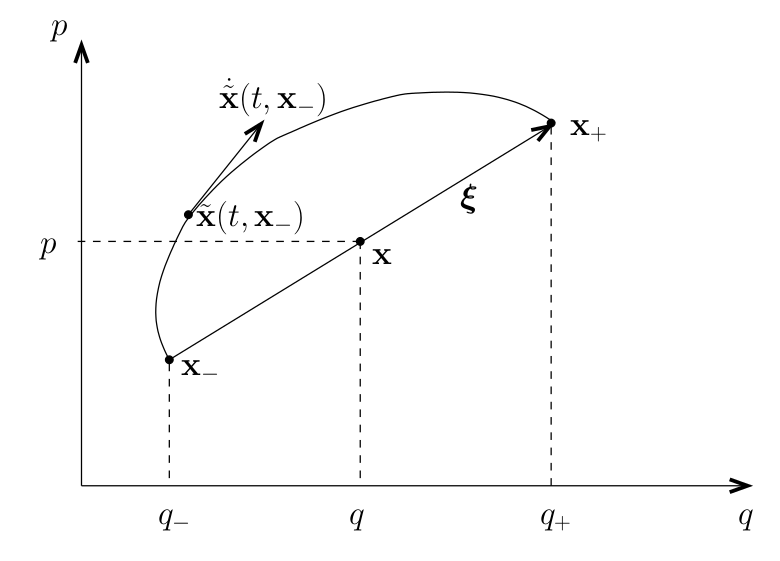}
\caption{The action, $S_t(\x)$, is just the symplectic area between
the trajectory from ${\x}^-$ to ${\x}^+$ and its chord $\Vxi$, from which one subtracts $Et$,
where $E$ is the energy of the trajectory.}
\label{Fig1}
\end{figure}

The difficulty is that the centre action is not generally expressed as a power series, so that
one cannot immediately continue analytically the SC expression $\e^{-\beta H}(\x) \approx U_{-i\hbar\beta}(\x)$,
in terms of the complex thermal time $t\mapsto -i\hbar\beta=-i\theta$, 
except in simple cases. In the short time limit, one can neglect the curvature of the trajectory, 
thus replacing the trajectory arc by the chord itself, whatever the Hamiltonian. In other words, 
one has $\Vxi \approx t \dot\x$, so that $S_t(\x) \approx -tH(\x)$ and $\M_t \approx \Id$. 
For the imaginary thermal time $t=-i\hbar\beta$, one then retrieves the classical high temperature limit 
of the thermal Wigner function: $W_\beta(\x) \approx P_{\beta}(\x)$. 

The Weyl propagator for the harmonic oscillator $(N=1)$ 
\be
H_h(x) = \frac{\omega}{2} ~(p^2 + q^2)
\label{HHO}
\ee
provides an illuminating example. After a time $t$ the Cayley matrix 
for this archetypical elliptic propagator will be $\mathbf{B}_t = - \tan(\omega t/2) \Id$ \cite{Report,OAI}, 
so that $S_t(\x)  = - \x \cdot \tan(\omega t/2) \x$ and the full Weyl propagator is
\be
U_t(\x) = \frac{1}{\cos(\omega t/2)} 
\exp \left[-\frac{i}{\hbar} \tan (\omega t /2)\; (p^2 + q^2) \right].
\label{Uho}
\ee
Here, the monodromy matrix, $\M_t$, does not depend on $\x$, leading to
the simple form taken by the square root in the amplitude of \eref{Uweyl}.
In general, the SC expressions for propagators generated by quadratic Hamiltonians
are exact and here it is easy to proceed to the analytic continuation for the imaginary thermal time,
$t=-i\theta$. Thus, the exact expression for the Weyl exponential of the Hamiltonian is
\be 
\e^{-\beta H_h}(\x) = \frac{1}{\cosh(\omega\theta/2)} 
\exp \left[-\frac{1}{\hbar} \tanh(\omega\theta/2)\;~ (p^2 + q^2) \right].
\label{Uiho}
\ee

One should observe that the Weyl propagator for the inverted harmonic oscillator, 
\be
H(x) = \frac{\omega}{2} ~(p^2 - q^2),
\label{invHO}
\ee
is
\be
U_t(\x) = \frac{1}{\cosh(\omega t/2)} 
\exp \left[-\frac{i}{\hbar} \tanh (\omega t /2)\; (p^2 - q^2) \right],
\label{Uinvho}
\ee
so that \eref{Uiho} is a curious blend of the expressions for the harmonic oscillator
and its inverse.
\footnote{Unlike the harmonic oscillator action, which is singular at $\omega t=\pi$, 
the action for its inverse stabilises smoothly due to the relevant trajectory merely hugging the stable
and the unstable manifold closer and closer  as $t$ increases. Due to structural stability, this
also holds for nonlinearly deformed unstable equilibria.} 
In this case, the Hamiltonian can be obtained from a Lagrangian.
Its thermal properties are then obtained from the real time propagation 
in the inverted oscillator  \cite{IngoldLec}.
Our treatment of general Hamiltonians in section 6 may be considered to be a further generalization, 
but first we deal here with a general quadratic Hamiltonian \eref{quadraticH}
and its quantization.

A crucial characteristic of the Weyl representation is its {\it symplectic invariance},
that is, the functions that represent operators in phase space are invariant
under linear canonical transformations of its arguments (see e.g. \cite{AlmHan82}).
Indeed these classical-like transformations are just a particularly apt expression of 
similarity transformations by the quantum {\it metaplectic group} of unitary operators
\cite{Littlejohn86,deGosson06,OAI}.
These are generated by the quantization of \eref{quadraticH}, such that the SC form is exact,
that is, for a single degree of freedom the Weyl propagators for the metaplectic operators are
 \be
U_t(\x|\mH) = \frac{1}{\cos(\Omega t/2)} 
\exp \left[-\frac{1}{\hbar} \tan (\Omega t/2)\; \frac{\x\cdot \mH~\x}{\Omega} \right],
\label{Uhogen0}
\ee
where $\Omega^2\equiv \det \mH$. 

Thus, the unitary metaplectic operators have the same general form as those which may evolve them
through a similarity transformation. This is not so for their analytic continuation as non-normalized thermal Wigner functions,
provided $\Omega^2>0$.
Rather, these {\it metaplectic Wigner} functions become the exact generalization of \eref{Uiho}:
\be
\e^{-\beta H}(\x|\mH) = \frac{1}{\cosh(\Omega\theta/2)} 
\exp \left[-\frac{1}{\hbar} \tanh (\Omega\theta /2)\; \frac{\x\cdot \mH~\x}{\Omega} \right]
\label{Uhogen}
\ee
with $\theta=\hbar\beta$. So one may include, for instance, a constant magnetic field
for a charged harmonic oscillator. Extending to $\beta<0$, 
the complete set of operators $\exp [-\beta~ \hat{\x} \cdot \mH \hat{\x}/2]$
form a group that is isomorphic to the metaplectic group of quantum flows.

In the case of a free particle, such that $\Omega\rightarrow 0$ as $\x\cdot \mH~\x \rightarrow {\vecp}^2/2m$,
one obtains
\be
\e^{-\beta H}(\x) = \exp \left[-\frac{\beta}{2m}~\vecp^2\right],
\ee
a non-normalized version of $P_{\beta}(\x)$. So one encounters the general rule
that the Weyl representation of any operator that is a function either exclusively of the momenta, 
or exclusively of the positions, has exactly its classical form.

For $N>1$ the transformation which diagonalizes the symmetric matrix $\mH$ 
in the positive quadratic Hamiltonian also places the matrix $\J\mH$ in its {\it Williamson normal form}
(see e.g. \cite{Arnold}), so that this transformation is symplectic. The only option of the normal form,
if the Hamiltonian is bounded from bellow, is that all degrees of freedom are elliptic.
Thus the thermal Wigner function at sufficiently low temperature reduces to a product 
of $N$ individual harmonic oscillator Wigner functions \eref{Uiho} with frequencies $\omega_n$,
each in its own phase plane.

The lowest correction to the action for a general Hamiltonian at short times is
\be
S_t(\x) \approx -tH(\x) - \frac{t^3}{24}~ \dot{\x} \wedge \ddot{\x}
=-tH(\x) - \frac{t^3}{24}~ \dot{\x}\cdot \mH_{\x}~\dot{\x}~;
\label{thirdorder}
\ee
the first version derived in \cite{Ber89b}, whereas the second in \cite{Report} 
evinces the connection with the local quadratic approximation of the Hamiltonian in terms of
its Hessian matrix $\mH_\x = \der^2 H(\x)/\der \x^2$. Indeed, following the appendix in \cite{Report} 
the local inhomogeneous quadratic approximation to the Hamiltonian
\begin{eqnarray}
H^{(2)}(\x'|\x) \equiv H(\x) + \vct{h_\x}\cdot (\x'-\x) + \frac{1}{2}~(\x'-\x)\cdot \mH_\x~(\x'-\x) ,
\label{localH1}			
\end{eqnarray}
where $\vct{h_\x}$ is the local gradient of the Hamiltonian, generates a symplectic flow in a kind of tangent space,
with points $(\x'-\x)$. 
We can now reexpress this in terms of the homogeneous quadratic Hamiltonian \eref{quadraticH} 
\be
H^{(2)}(\x'|\x) = [H(\x) -  H(\x - {\vct\gamma}_\x | \mH_\x)] +  H(\x' - {\vct\gamma}_\x | \mH_\x) ,
\label{localH2}
\ee
by defining the local {\it centre of curvature}: 
\be
{\vct\gamma}_\x \equiv \x -  {\mH_\x}^{-1}{{\vct h}_\x}. 
\ee
One should notice that \eref{localH2} holds even for $\x'\rightarrow \x$, 
so that one may evaluate the time derivatives $\dot{\x} = \J\mH_\x (\x-{\vct\gamma}_\x)$ 
and $\ddot{\x} = \J\mH_\x\J\mH_\x (\x-{\vct\gamma}_\x)$ at the centre $\x$ itself.
But for a single degree of freedom, $(\J\mH_\x)^2 = -\Omega_\x^2\Id$, 
so that the acceleration always points to the local centre of curvature:
$\ddot{\x}= -\Omega_\x^2 (\x-\vct\gamma_\x)$. 

The quantization of the approximate Hamiltonian \eref{localH2}, with $\x$ kept as a constant and further parametrized 
by the local centre of curvature $\vct\gamma_\x$, then identifies the correction to the short time
approximation for the centre action  \eref{thirdorder} as merely the third order
expansion in time of the action generated by a quadratic Hamiltonian: $\tan$ (for real time) and $\tanh$ 
(for imaginary thermal time). For a general Hamiltonian one thus obtains the short time
approximation of $\e^{-\beta H}(\x)$ as the generalization of the third order expansion 
of the action in \eref{Uhogen}, i.e.
\be
\hspace{-2,5cm}
\begin{array}{ll}
\e^{-\beta H}(\x)  &\approx \displaystyle
\frac{1}{1+\frac{(\hbar\beta\Omega_\x)^2}{8}}\; \exp{\left[- \beta H(\x) -  \frac{\hbar^2\beta^3}{24} \dot{\x}\cdot \mH_{\x}~\dot{\x} \right]} \\
 ~ & \approx\displaystyle\frac{1}{1+\frac{(\hbar\beta\Omega_\x/2)^2}{2}} 
\exp{\left[- \beta H(\x) 
-  \frac{(\hbar\beta\Omega_\x/2)^3}{3}\; 
\frac{(\x-\vct\gamma_\x)\cdot \mH_{\x}~(\x-\vct\gamma_\x)}{\hbar~\Omega_\x} \right]} .
\end{array}
\label{shorttW}
\ee
The second form is less convenient for calculations, since it requires the evaluation of the
local centre of curvature $\gamma_\x$, but it clearly evinces the dimensionless expansion parameter 
$(\hbar\beta\Omega_\x/2)$, where $\hbar \beta=\theta$ is the thermal time, while $\Omega_\x$ is the
angular frequency of trajectories within the local quadratic approximation.

A final step follows from the recognition that, to be consistent with the local quadratic approximation to the Hamiltonian, 
one can keep the whole series for $\tanh$ and $\cosh$ instead of just the third order in the dimensionless parameter.
This leads to a dampened version of the metaplectic Wigner function \eref{Uhogen} with the origin displaced to the
centre of curvature $\gamma_\x$:
\be
\e^{-\beta H}(\x)  \approx\e^{-\Delta(\x)} ~~ \e^{-\beta H}(\x-\gamma_\x|\mH_\x) . 
\label{metaW} 
\ee
The exponent of the attenuation factor in this {\it local metaplectic Wigner function}
is 
\be
\Delta(\x) = \beta (H(\x) - H(\x-{\vct\gamma}_\x | \mH_\x) = \beta (H^{(2)}(\gamma_\x|\x)),
\ee
so that quantum effects are only represented within the metaplectic Wigner function itself, 
the second factor in \eref{metaW}. One can then picture a tangent space at each point $\x$ as holding a
displaced generalized harmonic oscillator, from which the local metaplectic Wigner function is constructed. 
This alternative form then consists of a classical attenuation of a fully quantum thermal Wigner function,
replacing the quantum correction to the classical distribution obtained in (3.22).
It is not evident a priori, which of these approximations will best extend to increasing thermal time.
One should note that \eref{metaW} is nonsingular for all time, but its real time version will have the 
singularities of the Weyl propagator for the harmonic oscillator.

\section{Normal forms and the Kerr Hamiltonian}

For a temperature that includes energies beyond the range of validity of the quadratic approximation, 
the region surrounding the absolute minimum of the Hamiltonian can be transformed to its {\it Birkhoff normal form} 
\cite{Arnold,livro}, that is, for $N=1$,
\be
\fl H_{Bir}(\x) = \omega ~\left(\frac{p^2 + q^2}{2}\right) + H_2~  \left(\frac{p^2 + q^2}{2}\right)^2 
+ H_3~\left(\frac{p^2 + q^2}{2}\right)^3 + ... \equiv F\left(\frac{\x^2}{2}\right),
\label{nfH}
\ee
which simply separates into polynomials  $F_n\left(\frac{\x_n^2}{2}\right)$ for higher degrees of freedom.
Generally, for $N>1$, the infinite series has no hope of converging,
because the integrable normal form cannot cope with the intricate KAM-type motion that surrounds
the equillibrium, but a higher truncation than the simple quadratic can lead to a great improvement 
in the accuracy of trajectories for a finite time. Evidently this approximation cannot be extended
up to an eventual saddle point in the Hamiltonian, but the classic expansion by Gustavson \cite{Gustav}
for the Henon-Heiles Hamiltonian \cite{HenHei} (bellow the energy of its three saddles) 
supplies good trajectories for a finite time.

We cannot use the normal form transformation to transport exactly a Wigner function,
because this is a nonlinear transformation, even if it can be chosen to be canonical.
Nonetheless, this is just the {\it truncated Wigner approximation} \cite{Heller76} 
(TWA, also referred to as LSC-IVR \cite{Miller01}),
which is widely used and adequate if the Wigner function is not highly oscillatory.
At low enough temperatures for the spectral decomposition of the thermal Wigner function \eref{thermalW} 
to be of use, each term can be approximated by the normal form deformation
of $W_j(\x)$. Each of these is the product of eigenstates of the quantum Hamiltonians corresponding to  
$F_n\left(\frac{\x_n^2}{2}\right)$, with the partial energy for each degree of freedom 
approximated SC as $E_{nk} = F_n\left(\frac{1}{2}+k\hbar\right)$. 

Even if the energy is too high for individual states to be useful, 
the normal form may still extend smoothly the picture  of the motion 
beyond the quadratic approximation. Typically, for a Hamiltonian with a single parameter
the quadratic term only cancels at bifurcations of the equilibrium, calling for more
complex normal forms \cite{Arnold, livro}. However, the {\it Kerr Hamiltonian}
\be
H_K(\x)= [H_h(\x)]^2
\label{HKerr}
\ee
for the single photon Kerr effect \cite{Kirchetal}, i. e. simply the square of the harmonic oscillator,
can be treated as a special normal form. 
The quantum evolution of a Kerr system is not governed by an ordinary
second order Schr\"odinger equation, even though its quantum evolution is known exactly 
\cite{YuSto86,AvPer89} and it has been shown to be amenable to SC treatment \cite{Lanetal19}.

For simplicity, the discussion is now limited to a single (typical) degree of freedom. 
The classical trajectories generated by the normal form are identical to the arcs of concentric circles 
for the harmonic oscilator, but with the angular frequency of each arc depends on its initial value: 
$\omega(\x)= F'(\x^2/2)$, the derivative of $F$ with respect to its scalar argument. 
This allows one to calculate the action \eref{Uho} in a closed form, 
which can be continued analytically for complex thermal time, even though neither the SC Weyl propagator
nor the corresponding thermal Wigner function is exact.

Let us run forward and backward in time a pair of trajectories from a given point $\X$ for the duration $t/2$. 
Then the chord joining the endpoints of the full trajectory arc, spanning the angle $t F'(\X^2/2)$,
is centred on
\be
\x  = \cos\left[\frac{t}{2} F'\left(\frac{\X^2}{2}\right)\right]~~  \X.
\label{xX}
\ee
The area of the triangle formed by these endpoints with the origin is just $\sin[t F'(\X^2/2)]~ \X^2/2$,
which must be subtracted from the arc area $ t F'(\X^2/2)~\X^2/2$ to obtain the area 
between the trajectory arc and the chord. Finally, subtracting $t H(\X)= t F(\X^2/2)$, one obtains the implicit
expression for the action \eref{St1} as
\be
S_t(\x(\X)) = \left[t F'\left(\frac{\X^2}{2}\right) 
- \sin \left( t F'\left(\frac{\X^2}{2}\right)\right)\right]~\frac{\X^2}{2}
- t ~F\left(\frac{\X^2}{2}\right).
\label{SBirt}
\ee
The analytic continuation of the action for imaginary thermal time $-i\theta=-i\hbar\beta$ is then $S_{-i\theta}(\x(\X))=-i S_\theta^i(\x(\X))$ with
\be
S_\theta^i(\x(\X)) = \left[\theta F'\left(\frac{\X^2}{2}\right) 
- \sinh \left( \theta F'\left(\frac{\X^2}{2}\right)\right)\right]~\frac{\X^2}{2}
- \theta ~F\left(\frac{\X^2}{2}\right).
\label{SBirtheta}
\ee

The explicit expression obtained by substituting $\X$ by $\x$ is generally much messier,
but for the harmonic oscillator, one has simply $F' = \omega$, so that only the central term survives 
and so the actions become
\be
S_t(\x) = - \tan(t\omega/2)~ \x^2 ~~ {\rm and} ~~   S_\theta^i(\x) = - \tanh(\theta\omega/2)~ \x^2, 
\label{actionho}
\ee
in line with \eref{Uho} and \eref{Uiho}.

The amplitude of the SC Weyl propagator is determined by its action according to \eref{Uweyl2}, though it must be noted
that the derivatives in \eref{Hessian} are taken with respect to the centre $\x$ rather than $\X$.
The same procedure for the thermal action supplies the amplitude of the thermal Wigner function.
In practice, however, there is no advantage in working out the full explicit expression for the thermal action,
since the expectation of observables can be computed by integrating directly over the trajectory midpoint $\X$, 
as will be discussed in section 7. On the other hand, it is illuminating to consider the long thermal time limit
of $S_{\theta}^i(\x)$. Recalling the identity $ \sinh 2x = 2 \sinh x \cosh x$
and that, fixing $\x$, $\X \rightarrow 0$ exponentially with increasing thermal time according to \eref{xX}, 
one obtains in the limit a single term
\be
S_{\theta}^i(\x) \rightarrow - \tanh (\hbar\beta F'(0)/2) ~\x^2.
\ee
But $ F'(0) = \omega$, so that the SC approximation for the normal form collapses into the thermal Wigner function
for the harmonic oscillator just as discussed in the overview in section 2. The case of the Kerr Hamiltonian
is anomalous, since there is no limiting quadratic behaviour.

One should contrast this straightforward SC scenario of the thermal Wigner function 
for the normal form with the one for the Weyl propagator. The latter has singularities
in its action \eref{SBirt} and the amplitude of the SC propagator, 
i.e. {\it caustics}, which are not a feature of the exact quantum propagator.
They subdivide the phase space into distinct regions to be traversed with care,
as exemplified in the case of the Kerr Hamiltonian in \cite{Lanetal19}.
Thus, in comparison, the thermal Wigner function is remarkably well behaved for 
Hamiltonians with a single minimum.

Without claiming to have exhausted all possibilities for the analytic continuation of the classical
centre action, we henceforth embark on an alternative general approach. In passing, it should be recalled that
for ordinary Hamiltonians, with both the classical and the Weyl form
\be
H_s(\x) = \frac{\vecp^2}{2m} + V(\vecq),
\label{psquared}
\ee 
the classical dynamical system can be derived directly from a Lagrangian function. 
This allows for a remarkable simplification of the thermal density matrix in the position representation,
through the expedient of redefining the momenta:
Imaginary time entails an imaginary velocity in the Lagrangian \cite{IngoldLec} 
and hence one deals with {\it imaginary momentum} $\vecp^i$ in the Hamiltonian, 
such that  $i \vecp^i \equiv \vecp$, so that the only vestige of the complexification
is that $\vecp^2 \mapsto -(\vecp^i)^2$. 
\footnote{The switch ${\dot \vecq}^2 \mapsto -(\dot \vecq^i)^2$ in the Lagrangian is already a part
of Feyman's adaptation of the path integral for the density matrix, though the SC limit is not considered in \cite{Feyn72}.}
This turns stable equilibria into unstable points and vice versa,  
generalizing the outcome for the harmonic oscillator. 
Our purpose here is in a way to mimic such a procedure for the Lagrangian, 
but through a complexification of phase space variables variables for arbitrary Hamiltonians, 
so as to obtain a real general SC approximation for the thermal Wigner function.

\section{Double phase space}

The simple artifact of defining an imaginary momentum does not guarantee that
trajectories centred on arbitrary phase space points will be real, even for Hamiltonians of the form \eref{psquared}.
Nonetheless, the doubling of the phase space does allow for an analogous procedure for general Hamiltonians.
Let us first describe this for the Weyl propagator, by breaking up the quantum evolution operator into half-times:
\be
\e^{-it\hat{H}/\hbar} = \e^{-it\hat{H}/2\hbar}~\hat{\Id} ~\e^{-it\hat{H}/2\hbar}
= \e^{-it\hat{H}_+/2\hbar}~\hat{\Id} ~\e^{+it\hat{H}_-/2\hbar}
\ee
with $\hat{H}_\pm = \pm \hat{H}$. In this last form, one immediately recognizes the operator
$\hat{I}_L(t)$, which propagates the {\it quantum Loschmidt echo or the fidelity}, that is,
the overlap of two different evolutions of the same initial state \cite{ZamOA11}. 
In the present particular case the pair of quantum evolutions are merely the forward evolution 
in the interval $(0,t/2)$ and its time reversal $(0,-t/2)$ with the same Hamiltonian $\hat H$. 

The corresponding classical echo matches the outcome of the following procedure:  One picks a real initial value $\X$
and evolves this forwards and backwards in time, $0\leq t'\leq t/2$, forming the trajectories 
$\tilde{\x}_+ (\X,t')$, generated by $H_+$, and $\tilde{\x}_- (\X,t')$, generated by $H_-$. So one can picture the centre and the chord, 
defined in \eref{centran}, as having evolved continuously in time, 
from the initial conditions $\tilde{\x}(0)=\X$ and $\tilde{\Vxi}(0)=0$ respectively, so that
\be
\begin{cases}{
\tilde{\x}(t') =\frac{\tilde{\x}_+(t') + \tilde{\x}_-(t)}{2}~~,~~ 
\tilde{\Vxi}(t') = \tilde{\x}_+(t') - \tilde{\x}_-(t'), ~~{\rm or}~~ \\
 \tilde{\x}_\pm(t') = \tilde{\x}(t') \pm \frac{\tilde{\Vxi}(t')}{2}.} 
\end{cases}
\label{centrev}
\ee
Then one obtains the action in \eref{St1} as simply
\be
S_t(\x(\X)) = \int_0^{t/2} {\rm d}t' ~ \tilde{\Vxi}(t')\wedge \dot{\tilde{\x}}(t') - tH(\X),
\label{St2}
\ee

At this stage, it is advantageous to define a new variable $\y \equiv \J\Vxi$. This is the true conjugate
variable to $\x$ in the {\it double phase space} that arises in the semiclassical theory of open quantum systems 
\cite{BrOA10} and superoperators \cite{SarOA16}. In this expanded classical picture $\x=(\vecp, \vecq)$ stands for
a {\it double position}, whereas $\y=(\y_\vecp, \y_\vecq)$ assumes the role of its canonical {\it double momentum}.
Thus, the wedge product in \eref{St2} reduces to an ordinary dot product. Furthermore, it is now possible to define the {\it double Hamiltonian}
\be
I\!\!H(\x,\y) \equiv H_+(\x-\frac{\J\y}{2}) -  H_-(\x+\frac{\J\y}{2}) = H(\x_+) + H(\x_-)
\label{dH1}
\ee
such that Hamilton's equations in the enlarged phase space provide the appropriate equations of motion
for the centre and chord trajectories in \eref{centrev}:
\be
\begin{cases}{
\frac{\der I\!\!H}{\der \x} = \frac{\der H_+}{\der \x_+}-\frac{\der H_-}{\der \x_-}
= -\J(\dot{\x}_+ - \dot{\x}_-) = -\J\dot{\Vxi} = -\dot{\y} \\ 
\frac{\der I\!\!H}{\der \y} = \frac{\J}{2}\frac{\der H_+}{\der \x_+}+ \frac{\J}{2}\frac{\der H_-}{\der \x_-}
= \frac{\dot{\x}_+ + \dot{\x}_-}{2} = \dot{\x}.}
\end{cases}
\label{dHameq1}
\ee

For the initial point $\tilde\x_+(0)=\tilde\x_-(0)=\X$, one has $\dot{\Vxi}=\dot{\x}_+ - \dot{\x}_-=2\dot{\X}$,
whereas $\dot{\x} = (\dot{\x}_+ + \dot{\x}_-)/2 =0$ as depicted in Fig.2. Nonetheless, $\dot{\x}$
is only instantaneously zero, so that $\tilde{\x}(t')$ is not constant, 
because generally the curvature of the trajectory is nonzero. 
\footnote{In previous definitions of a double Hamiltonian \cite{OARiBR09, BrOA10}, $\y=0$ was an invariant plane, but not so here.}
In conclusion, the double Hamiltonian allows us to express the action
as a standard integral in terms of the double position $\x$ 
and its conjugate double momentum $\y$, evaluated on half the time interval:
\be
S_t(\x(\X)) = \int_0^{t/2} {\rm d}t' ~ \tilde{\y}(t')\cdot \dot{\tilde{\x}}(t') - \frac{t}{2} I\!\!H(\X).
\label{St3}
\ee

\begin{figure}
\centering
\includegraphics[width=.6\linewidth]{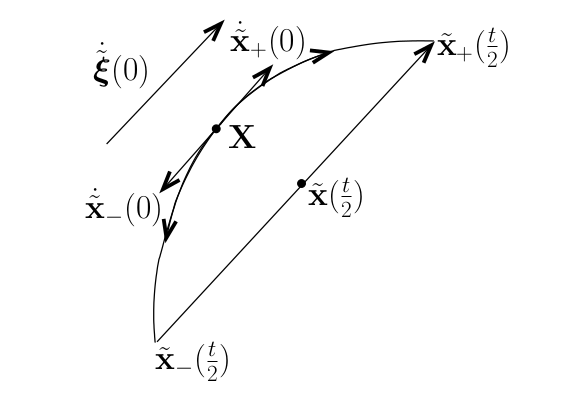}
\caption{The action for a doubled time, $S_{t}(\x(\X))$, is now built up from the forward and backward trajectories, 
with $\X$ as their initial value.}
\label{Fig2}
\end{figure}

The fact that the action is determined by the trajectory midpoint, $\X$, and is here dependent 
only indirectly on the centre, $\x$, may seem to be a disadvantage. On the contrary, it allows
us to dispense completely with the {\it root search} for trajectories in important applications.
Indeed differentiating the first equation in (5.2) leads to
\be
\frac{\der\tilde{\x}}{\der\X}(t/2) =\frac{1}{2} (\M_{t/2} + \M_{-t/2}),
\ee 
so that, recalling that $\M_{-t}=(\M_t)^{-1}$ and that $\det \M_t=1$ for all $t$,
the Jacobian determinant for the change of variables $\X \rightarrow \tilde{\x}(t/2)=\x$ is
\be
|\det \frac{\der\x}{\der\X}| = \frac{1}{2^N}~ |\det( \Id + \M_t)|.
\label{Jacobian}
\ee 
But this is just the square of the inverse amplitude of the Weyl propagator \eref{Uweyl}!

Let us consider the overlap of an initial function $|\psi\rangle$
with its evolution $|\psi_t\rangle = \hat{U}_t |\psi\rangle$. In terms of the corresponding 
Wigner function $W_\psi(\X)$, the exact expression is
\be
\langle \psi|\psi_t\rangle = {\rm tr}~ \hat{U}_t|\psi\rangle \langle \psi|
= \int {\rm d}\x~ W_\psi(\x)~U_t(\x).
\ee
Hence, the change of variable $\x \rightarrow \X$ leads to the SC approximation 
\be
\langle \psi|\psi_t\rangle \approx 
\int {\rm d}\X ~|\det \frac{\der\x}{\der\X}|^{1/2} \exp \left[ \frac{i}{\hbar}(S_t(\x(\X)))\right]
~ W_\psi(\x(\X))~.
\label{average2}
\ee
There is no extra integration and no worry about amplitude singularities at caustics, 
because these are cancelled by the new Jacobian factor in the integral.
In this way there is no search for trajectories, even though the Weyl propagator itself
is not an {\it initial value representation} (IVR), such as for instance the Herman-Kluk propagator 
\cite{HerKLuck,Miller01}. One should note that previous treatments of time evolution of overlaps
within the Weyl-Wigner representation \cite{IVRFVR} dealt only with its square-modulus,
that is, the self correlation of the evolved state.

\section{Complexified and decomplexified double phase space}

The complexification of time as $t\mapsto-i\hbar\beta$,
which has been adopted so far, leads to results that are equivalent to choosing the real {\it thermal time},
$\theta\equiv\hbar\beta$, together with the complex Hamiltonian
\be
H^i(\x)\equiv-iH(\x).
\label{Hi}
\ee
(This is imaginary for real $\x$, so that then $\dot \x$ is imaginary and all trajectories depart
from the real plane.) Such an option is verified to be legitimate in the short time limit
of the action \eref{thirdorder} and one can always build the actions for longer times
from the short time trajectory segments \cite{Report}.
For a general point in the complex phase space $\x \rightarrow \z$, Hamilton's equations
become
\be
\dot{\z} = \J\frac{\der H^i}{\der\z} = -i \J\frac{\der H}{\der\z}.
\ee
Evaluated at the complex conjugate point $\z^*$, one obtains $\dot{\z}(\z^*) =-[\dot{\z}(\z)]^*$,
which is the velocity for the reverse trajectory in time, leading in particular to an imaginary velocity whenever $\z=\z^*$.
Reiterating the process, $\ddot{\z}(0)$ will be real, and, more generally, 
the odd time derivatives will be imaginary whereas the even ones will be real. 
As a consequence, if we evolve the real phase space point $\X$ forwards and backwards in time, 
$0\leq \theta' \leq \theta/2$, forming the trajectories 
$\tilde{\z}_+(\X,\theta')=\z(0)+\theta' \dot{\z}(0) + \frac{(\theta')^2}{2}\ddot{\z}(0) + \ldots $ 
and $\tilde{\z}_- (\X,\theta')=\z(0)-\theta' \dot{\z}(0) + \frac{(\theta')^2}{2}\ddot{\z}(0) + \ldots$, 
then $\tilde{\z}_-(\X,\theta')=[\tilde{\z}_+(\X,\theta')]^*$ 
and one can define the evolving chord and centre:
\be
\begin{cases}{
\tilde{\x}(\X,\theta') = \frac{\tilde{\z}_+(\X,\theta') + \tilde{\z}_-(\X,\theta')}{2} = {\rm Re}~\tilde{\z}_+(\X,\theta')=\x^r \\  
\tilde{\Vxi}(\X,\theta') = \tilde{\z}_+(\X,\theta') - \tilde{\z}_-(\X,\theta') = 2i~{\rm Im}~\tilde{\z}_+(\X,\theta') = -i \Vxi^i,
}
\end{cases}
\ee
where $\x^r$ and $\Vxi^i$ are real.
Thus, one guarantees that the evolving centre trajectory remains real for all time, whereas the 
chord is always imaginary. But this is just what one should obtain from the complexification of the time
in \eref{centran}: the action being an odd function of time, it becomes purely imaginary for
imaginary time and the chord must follow suit.  

This preliminary remark will allow us to map the complexified double phase space into a real double phase space. Assuming analyticity of the action with respect to time, we extend the expression (\ref{St3}) to an imaginary time $t=-i\theta$, with $\theta \in I\!\!R$, and write
\be
S_{-i\theta}(\x(\X)) = \int_0^{-i\theta/2} {\rm d}t' ~ \tilde{\y}(t')\cdot \dot{\tilde{\x}}(t') - \frac{-i\theta}{2} I\!\!H(\X,0),
\label{St3i}
\ee
with the complex time generalization of (\ref{dHameq1}),
\be
\begin{cases}{
\y(0)=0 \\ 
\x(0) = \X \\ 
\frac{d \y}{d(-i\theta)} =  -\frac{\der I\!\!H}{\der \x}(\x,\y)  \\ 
\frac{d \x}{d(-i\theta)}  =  \frac{\der I\!\!H}{\der \y}(\x,\y).
}
\end{cases}
\label{dHameq_theta}
\ee
The preliminary remark suggests us to replace $\x$ by $\x^r$ and $\y$ by $-i\y^i$, leading to
\be
\begin{cases}{
\y^i(0)=0 \\ 
\x^r(0) = \X \\ 
\frac{d \y^i}{d\theta} =  -\frac{\der I\!\!H^r}{\der \x}(\x^r,\y^i)  \\ 
\frac{d \x^r}{d\theta}  =  \frac{\der I\!\!H^r}{\der \y^i}(\x^r,\y^i).
}
\end{cases}
\label{dHameq_real_theta}
\ee
with
\be
I\!\!H^r(\x^r,\y^i) = I\!\!H(\x,\y) = H(\x^r+i\frac{\J\y^i}{2}) + H(\x^r-i\frac{\J\y^i}{2}).
\label{dHr}
\ee
Notice that  $I\!\!H^r(\x^r,\y^i)$ is real if $\x^r$ and $\y^i$ are real, and therefore, since $\x^r$ and $\y^i$ are initially real, equations (\ref{dHameq_real_theta}) ensure that they will stay real for any $\theta$, thus confirming the preliminary remark. 

In its essence, the scenario depicted here is equivalent to our previous treatment of the
complex action \cite{BroMalAlm20}, but it is worth investigating it further. The full trajectory
in double phase space may be considered as combining both segments of $\tilde{\z}_\pm(\X,\theta')$ with $\theta'\leq \frac{\theta}{2}$, or, which is equivalent, the single phase space trajectory $\tilde{\z}(\z_-,\theta')$, with $\theta'\leq \theta$, generated by $H^i(\z)$, starting at $\z_-$ 
up to $\X$, and then on to its final point: $\tilde{\z}(\z_-,\theta)=\z_+$.
The real part of this trajectory is just $\x^r(\theta/2-\theta')$ from $\x(\X)$ to $\X$ and then 
it exactly retraces itself, whereas the imaginary part is $J\y^i(\theta/2-\theta')/2$ for $0\leq\theta'\leq \frac{\theta}{2}$ and then $-J\y^i(\theta'-\theta/2)/2$ for $\frac{\theta}{2}\leq \theta'\leq \theta$.
This simple behaviour is not the general rule for arbitrary initial points $\z_-$, 
so that it is an asset to be able always to start instead at the real midpoint, $\X$.

With these new real variables, one can see explicitly that the action (\ref{St3i}) is purely imaginary, that is
\be
S_\theta(\x(\X)) = -i S_\theta^i(\x(\X))
\label{St3theta}
\ee
where the real action $S_\theta^i(\x(\X))$ is defined by
\be
S_\theta^i(\x(\X)) = \int_0^{\theta/2} {\rm d}\theta' ~ \y^i(\theta')\cdot \dot{\x^r}(\theta') - \frac{\theta}{2} I\!\!H(\X,0). 
\label{Streal}
\ee
In other words, the actual exponent of the thermal Wigner function is $S_\theta^i(\x(\X))/\hbar$, and it is built on the dynamics of a trajectory in the real phase space $\left(\x^r,\y^i\right)$, that should be considered as a generalized position and a generalized momentum. Even so, it should be recalled that $I\!\!H^r(\X, 0)=2H(\X)$.
Viewed as a function of the centre $\x$, rather than the midpoint $\X$ of the full trajectory
generated by the double Hamiltonian $I\!\!H^r(\x^r, \y^i)$,
one should note that $S^i_{\theta}(\x)$ is exactly portrayed by \eref{St1}, so that this
trajectory satisfies the {\it centre variational principle} \cite{OA90,Report}:
In short, the action is stationary with respect to infinitesimal changes of the path in phase space,
which preserve its centre $\x$ (as opposed to its midpoint $\X$) . 

As an example to help overcome the unfamiliarity of these double phase space expressions,
let us return to the simple harmonic oscillator. Feeding its Hamiltonian
\eref{HHO} into the general expression for the double Hamiltonian \eref{dHr}, and leaving aside the labels $^r$ and $^i$ to aleviate notations, leads to
\be
\fl I\!\!H^r_h(\x,\y) = \omega~ \left(\x^2 - \Big(\frac{\y^2}{4}\Big)\right)
= - ~\omega~ \left(\Big(\frac{y_p^2}{4}\Big)-p^2  \right)- ~\omega~ \left(\Big(\frac{y_q^2}{4}\Big)-q^2  \right),
\label{dH5}
\ee
recalling that here one interprets the centre, $\x=(p,q)$, as a double phase space position,
whereas the (rotated) chord, $\y=(y_p,y_q)$, stands for the momentum. So we recognize that \eref{dH5}
is akin to the Hamiltonian for a double inverted oscillator \eref{invHO} (moving backwards). 
Resolution of (\ref{dHameq_real_theta}) gives $\x(\X,\theta) = \cosh(\omega\theta/2)~\X$ and brings the harmonic oscillator action \eref{actionho}.

The close connection established in section 2 between the thermal Wigner function 
for the harmonic oscillator and the third order short time approximation for general Hamiltonians
clearly implies that one can also generalize the procedure above, so as to obtain
this approximation within the double phase space approach. Indeed, it is interesting that the
original derivation in \cite{Ber89b} for the Weyl propagator already makes use of a construction 
around the midpoint of the trajectory at $t/2$, here labeled $\X$.

A further example of a double Hamiltonian is obtained for the Kerr system \eref{HKerr},
for which $I\!\!H^r_K(\x,\y)\neq [I\!\!H^r_h(\x,\y)]^2$. Indeed,
\be
\fl I\!\!H^r_K(\x,\y) = \frac{\omega^2}{2}~ 
\left[\Big(p^2 - \frac{y_p^2}{4}\Big)^2 + \Big(q^2 - \frac{y_q^2}{4}\Big)^2
+ 2\Big(p^2 - \frac{y_p^2}{4}\Big)\Big(p^2 - \frac{y_p^2}{4}\Big)-(qy_p-py_q)^2\right],
\label{dH6}
\ee
which has a nonquadratic equilibrium at the origin. In spite of the increased complexity,
note that $I\!\!H^r_K(\x,\y)=0$ along $\x =\pm\y$, so preserving something 
of the hyperbolic structure of $I\!\!H^r_h(\x,\y)$.

\section{Traces, averages and saddle points}

It is important to realize that the explicit action in terms of the  trajectory 
midpoint $\X$, instead of the centre $\x$ at which the thermal Wigner function is evaluated,
is no drawback for the evaluation of thermal averages. Indeed, the only difference with respect
to the previous case of the wave function overlap at the end of section 5 is that 
the real double Hamiltonian, which generates the classical trajectories for the real thermal time $\theta=\hbar\beta/2$ 
is expressed as \eref{dHr}, instead of \eref{dH1}. Hence, the same change of integration variable,
$\x=\tilde{\x}(\theta/2) \rightarrow \X$
with Jacobian \eref{Jacobian}, allows for the full semiclassical expression of the thermal average
of an arbitrary observable $\hat{O}$ to be expressed by a simple modification of \eref{average2}:
\be
\langle \hat{O} \rangle_\beta  \approx 
\frac{1}{Z_\beta}\int {\rm d}\X~|\det \frac{\der\x}{\der\X}|^{1/2} 
\exp \left[ \frac{1}{\hbar}(S^i_{\hbar\beta}(\x(\X)))\right]~ O(\x(\X)).
\label{exp3}
\ee

Here, the partition function $Z_\beta$ is evaluated by a similar integral in which $O(\x)=1$. 
For high temperatures corresponding to short imaginary times there will be no caustics, 
but even if these should arise, they would only
be zero curves, or more general zero  manifolds, of the Jacobian determinant. Should they arise, the residual task 
is then to evaluate an overall sign, which is the only remnant of {\it Maslov phases} 
in the Weyl propagator \cite{OAI} if caustics separate the domain of integration.

The expectation of functions of the energy itself $\langle F(\hat{H}) \rangle_\beta$ are important examples of the general
SC formula \eref{exp3}. They are related by standard thermodynamical relations, which are not quite obvious 
within the SC approximation, as we discuss in Appendix A.

The exact expression \eref{Uiho} supplies the partition function for the harmonic oscillator:
\be
\hspace{-2cm}
\begin{array}{lll}
Z_\beta &=& 
\displaystyle\int {\rm d}\x~\e^{-\beta H}(\x) \\ 
~ &=& \displaystyle\frac{1}{\cosh(\hbar\beta\omega/2)} 
\int {\rm d}\x~\exp \left[-\frac{1}{\hbar} \tanh(\hbar\beta\omega/2)\;~ (p^2 + q^2) \right]
= \frac{\pi\hbar}{\sinh(\hbar\beta\omega/2)}.
\end{array}
\label{partition2}
\ee
This diverges for $\beta \rightarrow 0$ and then decays
exponentially, i.e. much faster than in the classical high temperature approximation, 
for which $Z_\beta ={2\pi}/{\beta\omega}$.
Being that the exponent of the thermal Wigner function for the harmonic oscillator
is already quadratic, one may identify the above partition function with its own
saddle point approximation. For a general Hamiltonian, the validity of this approximation 
depends on large $\beta$ so as to confine the thermal Wigner function close to the saddle point at the origin, 
just the opposite of the high temperature approximation.
But this is just the condition for the normal form approximation for the thermal Wigner function 
in section 4 to hold, so that it merely collapses onto the central harmonic oscillator.

There is no saddle point contribution to the trace from periodic orbits in the
classical flow generated by the hyperbolic double Hamiltonian \eref{dH5}, except for
the equilibrium point at the origin itself. For the normal form approximation in section 4
the same is true. A saddle point evaluation of the expectation
of an observable, $\hat{O}$, according to \eref{exp3}, would be merely $O(0)$,
its Weyl value at the origin, which is obviously unsatisfactory at high temperatures,
for which the thermal Wigner function is not concentrated near the origin.

Thus, in the limit of low temperatures, one loses the complexity of the full semiclassical
theory of the thermal Wigner function when evaluating the thermal average of smooth observables.
It is required that the energy has a minimum for any thermal average even to be considered.
Then this minimum will correspond generically to an elliptic equilibrium, whatever
the number of degrees of freedom; the linearization of the flow 
(the quadratic approximation of the Hamiltonian) close to this origin
provides a (multidimensional) harmonic oscillator. The corresponding double Hamiltonian
will then be a (multidimensional) inverted harmonic oscillator, so that we
are guaranteed the absence of any periodic orbits in the saddle point approximation
of the trace, or the expectation of any smooth observable.

The special case of the Kerr Hamiltonian \eref{HKerr} reminds us that there may be no
quadratic approximation, but still there is a minimum. In this case, the high temperature 
approximation is 
\begin{eqnarray}
Z = \int {\rm d}\x~\exp \{- \beta[(\omega/2)\; (p^2 + q^2)]^2 \}
= \frac{\pi}{\omega}\sqrt\frac{\pi}{{\beta}}.
\end{eqnarray}
In contrast, the equilibrium of the quartic double Hamiltonian \eref{dH6} is not a minimum (or maximum), just as that of 
a hyperbolic Hamiltonian, so it is not surrounded by periodic orbits, which could contribute to the semiclassical
partition function.

\section{Discussion}

The density operator is the appropriate description of a system that is not isolated.
Indeed, a system described by an initially pure state, the projector corresponding to a vector in Hilbert space, 
is known to evolve into a mixed state, losing quantum coherence due to contact with an external environment.
Once the equilibrium of the system with the environment is achieved, it is characterized by the overall temperature 
$T=1/(\kappa_B\beta)$, so that the resulting mixed state can be identified with the one defined by the canonical ensemble, 
provided that the coupling to the environment is sufficiently weak. 

The observation that the product $\hbar \beta\equiv \theta$ has the dimension of time, a {\it thermal time},
permits an analogy of the static thermal density operator to the outcome of a {\it pseudo-evolution} in $\theta$. 
The semiclassical approximation of the thermal Wigner function breathes life into this metaphor, 
with the added quirk that one moves in thermal time from the initial classical canonical distribution, 
with no trace of Planck's constant, into the quantum realm. It should be emphasised that 
no other representation follows continuously the transition from the classical phase space distribution
to a full description of the thermal density operator.
 
Hamilton's equations and the entire classical pseudo-motion, on which the SC approximation is based,
are parametrized by thermal time. For small $\theta$, the classical trajectories are short,
so that the semiclassical approximation depends only on a local quadratic approximation
of the Hamiltonian. This provides the lowest quantum corrections to the high temperature limit.
Increasing thermal time requires the action of a longer trajectory, either immersed in a
complexified phase space \cite{BroMalAlm20}, or in the real doubled phase space presented here.
The fact that the appropriate double Hamiltonian has a saddle point instead of a stable equilibrium
prevents the relevant trajectory from growing beyond the constraint provided by the pair of 
stable and unstable manifolds. Thus its action is limited and converges for large $\theta$ 
to the exponent of the Wigner function of the dominant ground state in the low temperature limit. 

Having indulged in the metaphor of a {\it thermal-dynamical} system, we must reiterate
that only strictly equilibrium properties have been here considered. Some of our results
can be extricated from a previous paper on the complexified semiclassical approximation
of quantum work and its employment in the Jarzynski equality \cite{BroMalAlm20}.
But that was fundamentally a dynamical context, even if it may be considered in an adiabatic, quasi-statical limit.
So it is important to gain a full clear view of the rich underlying pure statics.
It should be pointed out that the present approximations are insensitive to any small effect of tunnelling
under saddle points on steady states, if the Hamiltonian has more than one minimum.

Digging deeper, one may even question the central assumption that the thermal equilibrium of the system can be independent 
of the (in most cases) uncontrollable environment and their coupling, when viewed in a full quantum scenario. 
After all, the system Hamiltonian does not in general  commute with the total Hamiltonian, which includes the
system, the environment and their coupling, so one is implicitly assuming the limit of weak coupling.
Indeed, the damping strength, characteristic of dissipative quantum evolution, will appear in the
equilibrium state. In the partition function, this dependence on the
damping strength is to be expected because it reflects the broadening of
discrete eigenstates \cite{Hanke1995,IngoldLNP}.

Notwithstanding the complexity of the general features of quantum equilibrium and the processes by which
it is attained, it is of fundamental importance to be able to deal with thermal quantum systems on their own,
described in their simplest form by the canonical density operator. The thermal Wigner function presents this
in a very convenient form, permitting the evaluation of the partition function and thermal averages
as {\it classical} phase space integrals. The semiclassical approximation to the thermal Wigner function 
bridges its extreme limits, while leading to simple quantum corrections to the classical high temperature limit. 
Further saddle point approximations would discard most of the information contained in the full semiclassical theory. 
On the other hand, full content is preserved by a mere shift of the integration variable 
of a thermal average, so dispensing with laborious searches for trajectories that are indirectly defined.
No restriction to systems that satisfy the ordinary second order Schr\"odinger equation impinges
on the present semiclassical approximation for the density operators of the canonical ensemble.

\appendix

\section{Thermodynamic relations}

The SC expectation of a smooth function of the Hamiltonian is a special case of \eref{exp3},
that is,
\be
\langle F(\hat{H}) \rangle_\beta  \approx 
\frac{1}{Z_\beta}\int {\rm d}\X~|\det \frac{\der\x}{\der\X}|^{1/2} 
\exp \left[ \frac{1}{\hbar}(S^i_{\hbar\beta}(\x(\X)))\right]~ F(H(\x)),
\label{exp4}
\ee
or equivalently the SC approximation to the thermal Wigner function is inserted directly in \eref{average}
to provide a macroscopic property of the system in complete thermal equillibrium with its environment.
In the case of the exact thermal average of the Hamiltonian itself the relation
\be
\langle \hat{H} \rangle_\beta = -\frac{1}{Z_\beta}\frac{d Z_\beta}{d\beta}
\label{firstid}
\ee
is widely employed, even though there is an implicit assumption of an evolution of the temperature
through the derivative, just as with the Jarzynski equality discussed in section 8. Then, the strength
of the coupling to the environment becomes relevant and indeed, one should restrict \eref{firstid} to the limit of weak coupling
\cite{Haenggi2006}. In any case it is easily seen to hold formally, whereas it is more problematic within the full SC approximation.

Even though \eref{firstid} is readily verified for a quadratic Hamiltonian (as it should, since here SC is exact),
there are two major difficulties for general Hamiltonians. The first is that the amplitude of the
Wigner function depends on $\beta$, not just its action. The amplitude cannot be taken out of the integral,
since it also depends on $\x$, so that, on taking the derivative within the integral, 
there is a second term beyond the derivative of the exponential. The second problem is that,
in taking the derivative of the exponent $S^r_{\hbar\beta}(\x(\X))$ itself, one obtains $H(\X)$ instead of $H(\x)$, 
together with the derivative of the integral in \eref{Streal}, which depends nonlinearly on $\beta$.
Thus, the approximate satisfaction of the relation \eref{firstid} depends on the counterbalancing 
of several terms and the direct evaluation of \eref{exp4} should be preferred to the RHS of \eref{firstid}.

Curiously, both approximations \eref{shorttW} and \eref{metaW}, which account for the lowest quantum correction
to the non-normalized thermal Wigner function, accommodate a version of the relation \eref{firstid}. 
Indeed, separating the exponent in the former into its linear and nonlinear parts,
\be
\fl \e^{-\beta H}(\x) \approx \frac{\exp \left[- \beta_1 H(\x)\right]}{1+(\hbar\beta_2 \Omega_\x/2)^2/2} 
\exp \left[-  \frac{(\hbar\beta_2\Omega_\x/2)^3}{3}\; 
\frac{(\x-\vct\gamma_\x)\cdot \mH_{\x}~(\x-\vct\gamma_\x)}{\hbar~\Omega_\x} \right] \Big{|}_{\beta_1 = \beta_2 = \beta},
\ee
one obtains a {\it double-$\beta$ partition function}
\be
\fl Z_{(\beta_1,\beta_2)} \equiv  \int d\x~\frac{\exp \left[- \beta_1 H(\x)\right]}{1+(\hbar\beta_2\Omega_\x/2)^2/2} 
\exp \left[-  \frac{(\hbar\beta_2\Omega_\x/2)^3}{3}\; 
\frac{(\x-\vct\gamma_\x)\cdot \mH_{\x}~(\x-\vct\gamma_\x)}{\hbar~\Omega_\x} \right] 
\ee
and thus a lopsided version of \eref{firstid}:
\be
\langle \hat{H} \rangle_\beta \approx -\frac{d \log Z_{(\beta_1,\beta_2)}}{d\beta_1}\Big{|}_{\beta_1 = \beta_2 = \beta}.
\label{secondid}
\ee
This same equality results from the isolation of the linear part of the exponent in the metaplectic Wigner function,
with the alternative definition of a double-$\beta$ partition function as
\be
\fl Z_{(\beta_1,\beta_2)} \equiv  \int d\x
\exp[-\beta_1 H(\x)] ~\exp[\beta_2 H(\x-{\vct\gamma}_\x | \mH_\x)] ~ \e^{-\beta_2 H}(\x-\gamma_\x|\mH_\x) .
\ee

Either of these definitions of a double-$\beta$ partition function will again introduce a first quantum correction
to further thermodynamic relations. The definition of the heat capacity again implies a dynamic (even if quasi-static)
process of feeding in heat and for finite coupling to the environment its definition is no longer unique
\cite{Haenggi2006}. However, the assumption of weak coupling to the enviroment allows to relate the heat capacity
to the second derivative of the partition function,
\be
\fl C \equiv \frac{d\langle \hat{H} \rangle_\beta}{dT}
= \frac{-1}{k_B T^2}\frac{d\langle \hat{H} \rangle_\beta} {d\beta}
=\frac{1}{k_B T^2}\left[\left(\frac{1}{Z_\beta}\frac{d Z_\beta}{d\beta}\right)^2 
-\left(\frac{1}{Z_\beta}\frac{d^2 Z_\beta}{d\beta^2}\right)\right]
=\frac{1}{k_B T^2} \Big[\langle \hat{H}^2 \rangle_\beta-\langle \hat{H} \rangle_\beta^2 \Big].
\label{thirdid}
\ee
The SC quantum correction at high temperatures is obtained as
\be
\fl C \approx \frac{1}{k_B T^2}\left[\left(\frac{1}{Z_{(\beta_1,\beta_2)}}\frac{d Z_{(\beta_1,\beta_2)}}{d\beta_1}\right)^2 
-\left(\frac{1}{Z_{(\beta_1,\beta_2)}}\frac{d^2 Z_{(\beta_1,\beta_2)}}{d\beta_1^2}\right)\right]_{\beta_1 = \beta_2 = \beta}.
\ee
Again, at lower temperatures where one needs a full SC approximation for the thermal Wigner function, the relation for
the specific heat is not transparently reproduced. Nevertheless, the variance of the energy, given by the final equality in \eref{thirdid}, 
is easily obtainable through the general formula for expectations \eref{exp3}, whilst recalling that the Weyl representation $H^2(\x)$ 
of $\hat{H}^2$ only equals the classical function $(H(\x))^2$, within a correction of order $\hbar$.

\section*{Acknowledgements}
We thank Raul Vallejos and Gabriel Lando for stimulating discussions.
Partial financial support from the 
National Institute for Science and Technology--Quantum Information
and CNPq (Brazilian agencies) is gratefully acknowledged.


\section*{Bibliography}
\begin{thebibliography}{99}

\bibitem{Feyn72} R.~P.~Feynman 1972
         \textsl{Statistical Mechanics} (Benjamin, Inc. Reading, Mass.)	
\bibitem{Gutzbook} M.~C. Gutzwiller 1990
				\textsl{Chaos in Classical and Quantum Mechanics} (Springer, New York)
\bibitem{IngoldLec} G-L. Ingold  2002
       in \textsl{Coherent Evolution in Noisy Environments} (eds. A. Buchleitner and K. Hornberger)
				(Springer, Berlin,LNP 611) 1							
\bibitem{Wigner} Wigner E~P 1932 Phys. Rev. {\bf 40} 749
\bibitem{Bertet02} Bertet P, Auffeves A, Maioli P, Ornaghi S, Meunier T,
                   Brune M, Raimond JM and Haroche S 2002 
         \emph{Phys. Rev. Lett.} {\bf 89} 200402
\bibitem{Grossmann} Grossmann A 1976 
         \emph{Commun. Math. Phys.} {\bf 48} 191
\bibitem{Royer} Royer A 1977 
         \emph{Phys. Rev. A} {\bf 15} 449	
\bibitem{Report} A.~M. Ozorio de Almeida 1998 
        \emph{Phys. Rep.} {\bf 295} 265.	
\bibitem{BroMalAlm20} O. Brodier, K. Mallick and A.~M. Ozorio de Almeida 2020
        \emph{J. Phys. A} {\bf 53} 325001					
\bibitem{OARiBR09} A. M. Ozorio de Almeida, P. M. Rios and O. Brodier 2009
        \emph{J. Phys. A}  {\bf 42} 065306.										
\bibitem{BrOA10} O. Brodier and A.M. Ozorio de Almeida 2010
        \emph{J. Phys. A} {\bf 43} 505308. 
\bibitem{Gronewold46} H.~J. Gr\"onewold  1946 
         \emph{Physica} {\bf 12} 405												
\bibitem{Arnold} Arnold VI  1978 
        \textsl{Mathematical Methods of Classical Mechanics} (Springer, Berlin)
\bibitem{livro} Ozorio de Almeida A~M 1988 
        \textsl{Hamiltonian Systems: Chaos and Quantization}
(Cambridge: Cambridge University Press)
\bibitem{Ber77} Berry M~V 1977 
        \emph{Phil. Trans. Roy. Soc A} {\bf 287} 237-71
\bibitem{AlmHan82} Ozorio de Almeida A~M  and J.~H.~Hanay 1982 
         \emph{Ann. Phys. NY} {\bf 138} 115.
\bibitem{Ber89b} M.~V. Berry 1989 
         \emph{Proc. R. Soc. Lond. A} {\bf 424} 279
\bibitem{Ber89} M.~V. Berry 1989 
         \emph{Proc. R. Soc. Lond. A} {\bf 423} 219	
\bibitem{Littlejohn86} R. G. Littlejohn 1986
       \emph{Phys. Rep.} {\bf 138} 193. 
\bibitem{deGosson06} M. de Gosson 2006
        \textsl{``Symplectic Geometry and Quantum Mechanics''} (Basel: Birkh{\"a}user Verlag)
\bibitem{OAI} A.~M. Ozorio de Almeida and G-L. Ingold 2014
        \emph{J. Phys. A} {\bf 47} 105303			
\bibitem{Gustav}F.~G. Gustavson 1966
       \emph{Astron. J.} {\bf 71} 670			
\bibitem{HenHei} M. Henon and C. Heiles, C 1964 
       \emph{Astron. J.} {\bf 69} 73								
\bibitem{Heller76} E.~J. Heller 1976
         \emph{J. Chem. Phys.} {\bf 65} 1289			
\bibitem{Miller01} W.~H. Miller 2001
         \emph{J. Phys. Chem. A} {\bf 105} 2942  							
\bibitem{Kirchetal} G. Kirchmair, B. Vlaskis, Z. Leghtas, S.~E.~Nigg, H.~Paik,
                    E. Ginossar, M.~Mirrahimi, L.~Frunzio, S.~M.~Girvin and R.~J.~Schoelkopf 2013				
         \emph{Nature (London)} {\bf 495}, 25 
\bibitem{YuSto86} B. Yurke and D. Stoler 1986
         \emph{Phys. Rev. Lett.} {\bf 57} 13
\bibitem{AvPer89} I.~S. Averbukh and N.~F. Perelman 1989
         \emph{Phys. Lett. A} {\bf 139} 449	
\bibitem{Lanetal19} G.~M. Lando, R.~O Vallejos, G.-L. Ingold and A.~M. Ozorio de Almeida 2019	
         \emph{Phys. Rev. A} {\bf 99} 042125						
\bibitem{ZamOA11} E. Zambrano and O.~A. Ozorio de Almeida 2011       
          \emph{Phys. Rev. E} {\bf 84} 045201	
\bibitem{SarOA16} M. Saraceno and A. M. Ozorio de Almeida 2016
				\emph{J. Phys. A} {\bf 49} 185302. 							
\bibitem{HerKLuck} M. F. Hermann and E. Kluk 1984 
        \emph{Chem. Phys.} {\bf 91} 27.			
\bibitem{IVRFVR} A.~M. Ozorio de Almeida, R.~O. Vallejos and E. Zambrano 2013
        \emph{J. Phys. A} {\bf 46} 135304 
\bibitem{OA90} A.~M. Ozorio de Almeida 1990
        \emph{Proc. R. Soc. Lond. A} {\bf 431} 403. 	
\bibitem{Lindblad} G. Lindblad  1976 
        \emph{Commun. Math. Phys.} {\bf 48} 119
\bibitem{Giulini} D. Giulini, E. Joos, C. Kiefer, J. Kupsch, I.-O. Stamatescu and H.~D. Zeh 1996
        \textsl{Decoherence and the Appearance of a Classical World in Quantum Theory} (Springer, Berlin)
\bibitem{DioKief} L. Diosi and C. Kiefer 2002 
        \emph{J. Phys. A} {\bf 35} 2675			
\bibitem{BroAlm04} O. Brodier and A.~M. Ozorio de Almeida 2004 
        \emph{Phys. Rev. E} {\bf 69} 016204
\bibitem{Hanke1995} A. Hanke and W. Zwerger 1995
	\emph{Phys. Rev. E} {\bf 52}, 6875.
\bibitem{IngoldLNP} G.-L. Ingold 2002
	\emph{Lect. Notes Phys.} {\bf 611}, 1
\bibitem{Haenggi2006}
	P. H{\"a}nggi and G.-L. Ingold 2006
	\emph{Acta Phys. Pol. B} {\bf 37}, 1537
 
          
\end{thebibliography}
\end{document}